\documentclass{article}

\usepackage{graphicx}
\usepackage{amsmath,amssymb,amsfonts}
\usepackage{color}
\usepackage{xr}

\renewcommand{\d}{\mathrm{d}}

\definecolor{changes}{rgb}{0,0,1}

\newcommand{\non}{\nonumber}
\newcommand{\mean}[1]{\langle #1\rangle}
\newcommand{\pbr}[1]{\left( #1 \right)}
\newcommand{\bbr}[1]{\left[ #1 \right]}
\newcommand{\cbr}[1]{\left\{ #1 \right\}}

\newcommand{\abs}[1]{\left| #1 \right|}

\newcommand{\norm}[1]{\left\Vert #1 \right\Vert}
\renewcommand{\d}{\mathrm{d}}

\newcommand{\order}[1]{\mathcal{O}\left( #1 \right)}
\DeclareMathOperator*{\argmax}{argmax}
\DeclareMathOperator*{\argmin}{argmin}
\DeclareMathOperator*{\ExpEi}{ExpEi}

\begin{document}
{\flushleft\Large\textbf{The value of monitoring to control evolving populations}}\\[1ex]

Andrej Fischer$^{1}$,
Ignacio Vazquez-Garcia$^{1,2}$ and
Ville Mustonen$^{1,3}$\\[1ex]

{\flushleft
${}^1$Wellcome Trust Sanger Institute, Hinxton, Cambridge, CB10 1SA, UK\\[1ex]
${}^2$Department of Applied Mathematics and Theoretical Physics, Centre for Mathematical Sciences, University of Cambridge, Wilberforce Road, Cambridge CB3 0WA, UK
${}^3$ E-mail: vm5@sanger.ac.uk
}

\section{Abstract}
Populations can evolve in order to adapt to external changes. The capacity to evolve and adapt makes successful treatment of infectious diseases and cancer difficult. Indeed, therapy resistance has quickly become a key challenge for global health. Therefore, ideas of how to control evolving populations in order to overcome this threat are valuable. Here we use the mathematical concepts of stochastic optimal control to study what is needed to control evolving populations. Following established routes to calculate control strategies, we first study how a polymorphism can be maintained in a finite population by adaptively tuning selection. We then introduce a minimal model of drug resistance in a stochastically evolving cancer cell population and compute adaptive therapies, where decisions are based on monitoring the response of the tumor, which can outperform established therapy paradigms. For both case studies, we demonstrate the importance of high-resolution monitoring of the target population in order to achieve a given control objective: to control one must monitor.

\begin{itemize}
\item{stochastic optimal control, adaptive cancer therapy, decision-making under uncertainty}
\end{itemize}

\section{Introduction}
The progression of cancer is an evolutionary process of cells driven by genetic alterations and strong selective forces \cite{Stratton:2009,Yates:2012}. The continued failure of cancer therapies to significantly reduce mortality, despite a host of new targeted cancer drugs, is largely caused by the emergence of drug resistance~\cite{Gillies:2012}. Cancer therapy faces a real dilemma: the more effective a new treatment is at killing cancerous cells, the more selective pressure it provides for those cells resistant to the drug to take over the cancer population in a process called \emph{competitive release}~\cite{Wargo:2007,Greaves:2012p4806}. \\

A genetic innovation conferring resistance 
can either be already present as standing variation or in close evolutionary reach, via \emph{de novo} mutations. The probability of these events is directly proportional to the genetic diversity of the tumor. Therefore, resistance is a problem especially for genetically heterogenous cancers~\cite{Gerlinger:2010p4474}. This diversity can be the result of a variable microenvironment, with different pockets of acidity, blood supply and geometrical constraints of surrounding tissue~\cite{Gillies:2012}. 
Also, late stage cancers not only carry the cumulative archaeological record of their evolutionary history~\cite{NikZainal:2012} but can also become genetically unstable and fall victim to
chromothripsis~\cite{Forment:2012}, \emph{kataegis}~\cite{NikZainal:2012b} and other disruptive mutational processes~\cite{Alexandrov:2013,Fischer:2013}. Thus, the probability of treatment success is higher in genetically homogenous and/or early stage cancers~\cite{Bozic:2012}. Taken together, these considerations place emphasis on early detection of tumors.\\
In cases where early detection is not achieved, the pertinent question is how to avoid treatment failure in the presence of genetic heterogeneity, which seems to be the norm for most solid cancers. One obvious attempt is to make treatments more complex and thus put the resistance mechanisms out of reach of the tumor. In combination therapy, the tumor is simultaneously treated with two or more drugs that would require different, possibly mutually exclusive, escape mechanisms for cells to become resistant. This approach has proven to be very successful in the treatment of HIV, where drug combinations are increasingly chosen based on genetic screens of a patient's virus for exploitable mutations~\cite{Gulick:1997,Hammer:1997,Lengauer:2006,Bock:2012}. In the context of cancer, this form of personalized therapy is not yet widely realized, mainly because of the much richer repertoire of genetic variation and adaptability of cancer cells and a comparable shortage of  drugs targeting \emph{distinct} biological pathways. For a recent study of the conditions under which combination therapy is expected to be successful in cancer, see~\cite{Bozic:2013p5240}. \\
For application of single drugs, there are a number of studies that concentrate on how the therapeutic protocol itself can be optimized.
 
It was realized that \emph{all-out} maximum tolerated dose chemotherapy is not the only, or necessarily the best, treatment strategy~\cite{Read:2011}.  Alternative dosing schedules were proposed such as drug holidays, metronome therapy~\cite{Foo:2009} and  \emph{adaptive} therapy~\cite{Gatenby:2009}. The realization of Gatenby et al. in \cite{Gatenby:2009} is that cancer, as a dynamic evolutionary process, can be better controlled by dynamically changing the therapy, \emph{depending on the response of the tumor}. Their protocol of reducing the dose while the tumor shrinks and increasing it under tumor growth showed a drastic improvement of life expectancy in mice models of ovarian cancer~\cite{Gatenby:2009}. Furthermore, Gatenby et al. made the important conceptual step of reformulating cancer therapy to be not necessarily about tumor eradication, but instead a dynamical problem where maintenance of a stable tumor size can be preferable. \\
Motivated by this experiment, we conjecture that there are substantial therapy gains in optimal applications of existing drugs, as of yet under-exploited. As a first step towards utilizing this potential we would like to formalize the intuition of Gatenby et al. To this extent, we aim to establish a theoretical framework for the adaptive control of evolving populations. In particular, we connect the idea of adaptive therapy to the paradigm of stochastic optimal control, also known as a Markov decision problem. For other applications of stochastic control in the context of evolution by natural selection see~\cite{Rivoire:2011,Rivoire:2014}. This is a well-established field of research which provides not only a natural language for framing the task of cancer therapy, but also a set of general purpose techniques to compute an \emph{optimal} control or therapy regimen for a given dynamical system and a given objective, such as population size reduction. While we demonstrate the main steps in this program,

we focus on the detrimental effect of imperfect information and the loss of control it entails. Our key conceptual result can be summarized as: \textit{to control, one must monitor}.\\
We first introduce the concepts of stochastic optimal control using a minimal, but non-trivial evolutionary example: how to keep a finite population polymorphic under Wright-Fisher evolution by influencing the selective difference between two alleles? 
If \emph{perfect} information about the population is available, the polymorphism can be maintained for a very long time. We will show how \emph{imperfect} information due to finite monitoring can lead to a quick loss of control and how some of it can be partially reclaimed by informed \emph{pre-emptive} control strategies. We then move to our main problem and introduce a minimal stochastic model of drug resistance in cancer that incorporates features such as variable population size, drug sensitive and resistant cells, a carrying capacity, mutation, selection and genetic drift. After computing the optimal control strategies for a few important settings under perfect information, we demonstrate the effect of 
imperfect monitoring. If only the total tumor size can be monitored, we show how a control strategy emerges that can adaptively infer, and thus exploit, the inner tumor composition of susceptible and resistant cells.

\section{Controlling evolving populations}

One can think about cancer therapy as the attempt to control an evolving population by means of drug treatment. Usually, the drug changes some of the parameters of the evolutionary process, such as the death rate of drug-sensitive cells. With application of the drug, one can thus actively influence the dynamics of the stochastic process and change its direction. All this happens with a concrete aim, such as to minimize the total tumor burden \emph{in the long term}. To introduce some of the concepts of stochastic optimal control, we use an example with a non-trivial control task.\\
 Imagine a bi-allelic and initially polymorphic population of constant size $N$ under the Wright-Fisher model of evolution~\cite{Ewens:2004}, i.e. binomial re-sampling of the population in each generation. The $A$ allele confers a selective fitness advantage of size $\sigma = N\,(f_A-f_B) \gg 1$ over the $B$ allele and will, without intervention, eventually take over the entire population (see Figure~\ref{Fig:WFcontrol}A). Assuming mutation to be negligible, the task at hand is to avoid, or at least delay, such a loss in diversity.
Now assume that we can change the selection coefficient externally by a quantity $u \in \bbr{-u_c,0}$ in the form $\sigma\to \sigma + u$. Ideally, we would have $\abs{u_c} > \sigma$, but this is not a necessary condition. 
In this setting, the control problem can be stated as follows: for a population at initial frequency $x_A(0) = x_0 = n_0/N \in\pbr{0,1}$, what is the optimal control protocol $\bar{u}_{0:T}(x_0)$ that maximizes the probability $P_a(T,x_0)$ that the polymorphism is still alive after a time $T$? 
\begin{gather}
\bar{u}_{0:T}(x_0) = \argmax_{u_{0:T}}\ P_a(T,x_0),
\label{Eq:u1}
\end{gather}
where the maximization is to be done over all possible sequences of controls $u_{0:T} = u_0 u_1 \dots u_T$ ($u_t\in \bbr{-u_c,0}$). Note that the stochastic nature of the process makes the control optimal only in the sense of expected outcomes. Individual realizations might well fall short of, or exceed, the implied mean survival time. 

It is helpful to picture a large ensemble of such populations, all starting off at frequency $x_0$, all being individually nudged by selection according to a (yet to be found) optimal protocol. When a trajectory hits one of the boundaries before the final time $T$, it is lost. The optimal control can thus also be seen to minimize this cost of attrition.
The standard technique to solve problems of this kind is to use a dynamic programming ansatz. Assuming that the partial problem 
for some intermediate starting point $(x_t,t)$ ($0<t<T$) has already been solved, we define the \emph{cost-to-go} $J(x_t,t)$ as the expected cost to be paid starting from $x_t$ at time $t$.
From this definition follows a backward-recurrence relation for $J$: the cost-to-go at ($x_t,t$) is the cost-to-go one time step later, but averaged over all possible states at that time. Which states most contribute to that average depends, via the propagator $W$, on the control $u_t$ one applies now~\cite{Kappen:2005}:
\begin{gather}
J(x_t,t) = \min_{u_t}\ \sum_{x'} J(x',t+1)\ W(x' \mid x_t; u_t),\label{Eq:C2G1}
\end{gather}
The absorbing boundary conditions take the form $J(0,t)=J(1,t) = T-t$. The right hand side of eq.~\ref{Eq:C2G1} can also include a term $V(x_t,u_t)$ that describes the \emph{potential} cost to be at $x$ and the \emph{control} cost to apply $u$. Here, both are assumed zero and cost is paid only at the boundaries. But it is important to keep in mind that optimal control problems are usually about achieving a certain goal with the least expense. For Wright-Fisher evolution, the transition matrix $W$ can be expressed as the probability under binomial sampling to draw $n_A' = N\, x'$ individuals of the $A$ allele. The crucial computational advantage of this relation is that the hard optimization in the space of all control protocols $u_{0:T}$ is exchanged for a  simple scalar optimization over $u_t$. In statistical physics, this technique is referred to as transfer-matrix method.
The intuitive interpretation of eq. \ref{Eq:C2G1} is that the decision for a control \emph{now} relies on  \emph{future} controls to be carried out optimally. In practice, the results of the local optimizations ($\bar{u}(x_t,t) = \argmin_{u_t}\dots$) constitute the optimal control to be applied when the system is at $x_t$ at time $t$. In many applications~\cite{Kappen:2005}, it is also useful to consider a \emph{receding} time horizon, such that $\bar{u}(x,0)$ is a stationary control.\\
The infinitesimal form of eq.~\ref{Eq:C2G1} in the diffusion approximation ($N\to\infty$, while $\sigma,u_c$ fixed and $\tau=t/N$) is called the Hamilton-Jacobi-Bellmann (HJB) equation~\cite{Bellman:1964,Kappen:2005},
\begin{gather}
-\partial_{\tau} J(x,\tau) = \min_{u_{\tau}}\ x(1-x)\bbr{ (\sigma+u_{\tau})\, \partial_x + \tfrac{1}{2}  \partial_x^{2}} J(x,\tau),
\label{Eq:HJB}
\end{gather}
together with the boundary condition for eq.~\ref{Eq:C2G1}.
Instead of attempting a direct solution of the HJB eq.~\ref{Eq:HJB} for the Wright-Fisher example, we will guess the solution and confirm it by direct numerical application of eq.~\ref{Eq:C2G1} (see also Figure~\ref{Fig:xc} in Supporting Information).

\subsection{Optimal control of a Wright-Fisher population with perfect monitoring} 
The optimal control function $\bar{u}(x_t,t)$ maximizes the probability that a polymorphism is still present after a time $T$. In the infinite horizon time limit $T\to\infty$, where the optimal control becomes stationary $\bar{u}(x)$, we expect it to also maximize the mean first passage time $\mean{T}_x$ out of the interval $0<x<1$ (for any x). Because $u$ appears linearly in eq.~\ref{Eq:HJB}, it is clear that only the two extreme control strengths are ever used to steer the system. This particular type of control (when control itself is free) is called \emph{bang-bang}~\cite{Kappen:2005}. It follows that the control profile $\bar{u}(x)$ will have the form of a step-function with critical frequency $x_c$,
\begin{gather}
\bar{u}(x) = \begin{cases}
0, &x<x_c\\
-u_c, &x\geq x_c
\end{cases}\
\Rightarrow\ \sigma + \bar{u} = \begin{cases}
 > 0, &x<x_c\\
 < 0, &x\geq x_c
\end{cases}.
\end{gather}
The only remaining parameter is the critical threshold $x_c(\sigma,u_c)$. To find an expression for the objective function $T_a(x_c,\sigma,u_c)$, we can consider the optimally controlled system as the simplest example of evolution under frequency-dependent, piecewise constant selection $\sigma+\bar{u}(x)$. The mean first passage time can be found analytically using standard methods for stochastic processes~\cite{Gardiner:2009} (see Supporting Information and Figure~\ref{Fig:xc}). At the correct threshold and with strong selective forces ($\sigma,\abs{\sigma+u_c}\gg1$), the gains are substantial and the polymorphism can be maintained for very long times.

\subsection{Loss of control due to imperfect monitoring}
The main assumption made so far was that \emph{perfect} information is available about the state of the system in the form of continuous (in time), synchronous (without delay) and exact (without error) measurements of $x$. These requirements are impossible to achieve in practice, when monitoring is always imperfect. 
As we will see, when the assumption of perfect information is relaxed, not only is control over the system lost, but the control profile $\bar{u}(x)$ also ceases to be \emph{optimal}. Rather than turning to the theory of partially observable Markov decision problems 
~\cite{Cassandra:1994}, we will use numerical analysis to demonstrate the effect of monitoring with finite resolution in time (relaxing the first condition).\\
Consider the situation where measurements of the frequency $x$ are given only at discrete times $\cbr{\tau_i}$, while no information is available during the intervals of length $\Delta = \tau_{i+1}-\tau_i$. The immediate question is: given a measurement $x_i$, what control should one apply while waiting for the next measurement? The perfect-information control $\bar{u}(x_i)$ is correct only \emph{initially}, and thus only in the limit $\Delta\to0$. But it is intuitively clear that a naive protocol, applying $\bar{u}(x_0)$ during the entire interval $\Delta$, cannot be optimal, because it does not anticipate the dynamics of $x$ under this regime (see the decrease in survival time in Figure~\ref{Fig:WFcontrol}B). For example, for $0<x_i<x_c$, the initial control is $\bar{u} = 0$ and the frequency will, on average, increase and eventually cross the threshold $x_c$. If one could observe the population at that point, the control should be switched to $\bar{u}=-u_c$ until $x$ crosses $x_c$ from above. The total result of the naive strategy is to amplify fluctuations due to this over-shooting.

\subsection{Playing-to-win vs. playing-not-to-lose}
\begin{figure}
\begin{center}
\includegraphics[width=8.7cm]{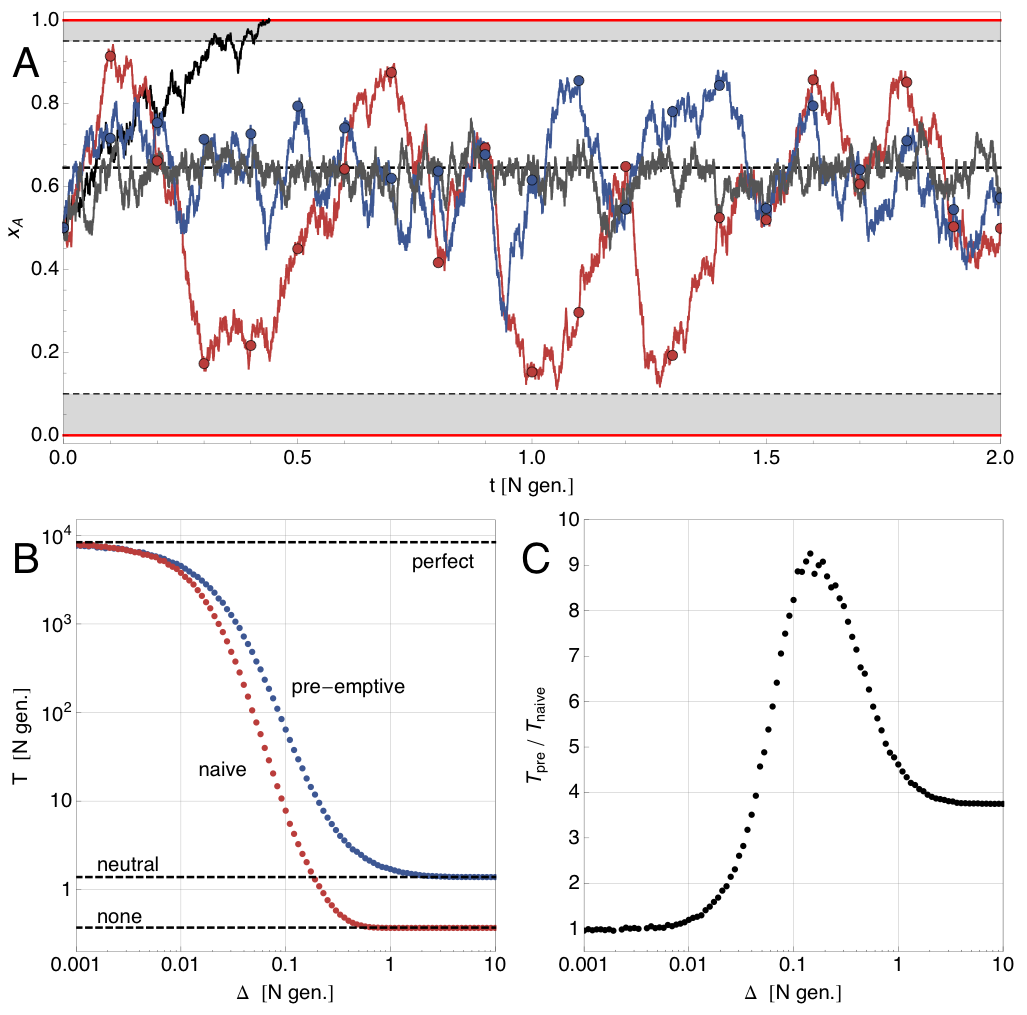}
\caption{Optimal control of a finite population under Wright-Fisher evolution in order to maintain an initial polymorphism. The intrinsic selection coefficient is $\sigma=10$ and control shifts selection to $\sigma+u$. 
(A) Sample trajectories starting at $x_0=0.5$: without control ($u\equiv0$, black line) the polymorphism is lost on a time scale of $1/\sigma$. With optimal control under perfect information (gray line, $\sigma+u=10$ for $x<x_c\approx0.644$, else $\sigma+u=-20$), it can be maintained for an average of $8000$ $N$ generations. With finite monitoring ($\Delta=0.1$, measurements $x_i$ at circles), naive control ($u\equiv\bar{u}(x_i)$), red line) is prone to over-shooting, while pre-emptive control (blue line) tries to avoid this by switching to a neutral regime after a certain time. 
(B) Loss of control under finite monitoring: as $\Delta$ grows, so does the probability that the polymorphism is already lost at the next measurement. Shown is the mean survival time over $5000$ trajectories with $N=10^4$.
(C) Under pre-emptive control, some of the loss of control can be regained, especially for intermediate values of $\Delta$.
}
\label{Fig:WFcontrol}
\end{center}
\end{figure}

Without a continuous flow of observations as input, a \emph{pre-emptive} control protocol $u^{\ast}(\tau,x_i)$ during the interval $\Delta$ must be pre-computed and then faithfully carried out. In the discrete-time (Wright-Fisher evolution) setting, there are $N\Delta$ generation updates until the next measurement and therefore $2^{N\Delta}$ different protocols to choose from. However, the example above suggests to search for the pre-emptive control in a much smaller space, namely within those protocols that start with either $u^{\ast}=0$ or $u^{\ast}=-u_c$ and then switch, at some later time $\tau_c(x_i,\Delta)$, to a \emph{neutral} regime with $u^{\ast}=-\sigma$ (the complexity of this space is only $2\,N \Delta$). The effect of such a control scheme is to move the population to a safe place and then try to keep it there. There are two important observations: first, this informed control outperforms the naive protocol significantly, especially for intermediate values of $\Delta$ (see Figure~\ref{Fig:WFcontrol}B-C); second, the safe parking position moves away from the boundary towards $x=0.5$ for larger values of $\Delta$ (see Figure~\ref{Fig:win-not-lose}). This shift from an aggressive control strategy under perfect information ($\Delta=0$, $x_c$ close to a boundary) to a more and more conservative one (aiming for $x=0.5$ and trying to stay there) can be summarized as \textit{playing-to-win vs. playing-not-to-lose}.\\
A similar loss of control can be expected for other types of monitoring imperfections and is a general feature of stochastic optimal control. It is important to note that the perfect-information control problem, and its solution $\bar{u}$, is a necessary starting point for the analysis. The naive control protocol above is indeed optimal for $\Delta\to0$, and still a very good option for $\Delta\ll 1$. In most cases, as we will see in the adaptive cancer therapy model below, finding $\bar{u}$ is challenging in itself and can be a good guidance for finding well performing control protocols even under imperfect conditions.

\section{Application to adaptive cancer therapy}
With the example above -- how to control a population aiming to maintain a polymorphism -- we introduced some key elements of stochastic optimal control and the basic steps of such an analysis. Here we apply these ideas to the problem of adaptive cancer therapy. We first introduce a minimal stochastic model of drug resistance in cancer. For different qualitative regimens, we then find the optimal adaptive therapy with perfect information. Finally, we extend these ideas to the case, where only the total cell population size can be observed but no readout of the fractions of susceptible and resistant cells is available.

\subsection{A minimal model of drug resistance in cancer} 
The desired features of a minimal model of drug resistance in cancer include: (i) a variable tumor cell population size $N$, (ii) at least two cell types, drug-sensitive and drug-resistant, (iii) a carrying capacity $K$ that describes a (temporary) state of tumor homeostasis, (iv) the possibility for mutation and selection between the cell types. Control over the tumor can be applied via a drug that changes the evolutionary dynamics by increasing, for example, the death rate of sensitive cells. We will assume here, as others have done in the context of cancer~\cite{Bozic:2012}, a well mixed cell population where the birth (or rather duplication) rate of cells is regulated by the carrying capacity. The dynamics of the model we have chosen here is encapsulated in the following birth and death rates for sensitive and resistant cells,
\begin{align}
B_i(n_s,n_r) &= \frac{(1+g_i)\, n_i}{1 + g \frac{N}{K} + s \frac{n_s}{K}} + \mu_0 (n_{\bar{i}}-n_i),
\label{Eq:BD1}\\[1ex]
D_i(n_s,n_r) &= n_i\,(1+ F_i(u)),\quad i\in\cbr{s,r},\, \bar{s}=r,\, \bar{r}=s\non
\end{align}
where $g_s=g+s$, $g_r=g$, $\mu_0$ is the mutation rate between cell types and $F_i$ encodes the effect of the drug ($u=1$) or its absence ($u=0$) on cell type $i$.  For $N\ll K$, the absolute growth rates are $g_i - F_i$. A drug effect of the form 
\begin{gather}
F_s(u) = u f_s\quad {\rm and}\quad F_r(u) = (1-u)f_r,
\label{Eq:DrugEff}
\end{gather}
renders the drug effective if $f_s>g+s$. The value $f_r=0$ corresponds to drug-resistance as such, but $f_r>g$ implies that resistant cells thrive under the drug and are drug-\emph{addicted}. 
Such an effect has been observed in mice with \textit{BRAF}-mutated melanoma treated with vemurafenib~\cite{DasThakur:2013}. Altogether, sensitive and resistant cells initially grow  exponentially until the total population size $N=n_s+n_r \approx K$. At that stage, competition for resources, space etc. becomes fierce. If sensitive cells have a differential growth advantage $s>0$ (they might not have to maintain an expensive resistance mechanism), resistant cells will eventually be removed from the tumor or reduced to a small  fraction (of size $\mu_0/s$). In reality, this scenario might not materialize, as the next mutation could propel the tumor into a new phase of exponential growth.\\
For the stochastic version of this process we can assume independent and individual birth and death events with the above probabilities per unit time. In analogy to the Wright-Fisher binomial update rule, here we can use a Poisson-like update.
\begin{gather}
n_i \to n'_i = n_i+\Delta n_i,\ {\rm with}\ \Delta n_i = \Delta n_i^+ - \Delta n_i^-
\label{Eq:BD2}\\
 \Delta n_i^+ \sim {\rm Pois}(B_i),\ \Delta n_i^- \sim {\rm Pois}(D_i),\ \mean{\Delta n_i} = B_i - D_i\non
\end{gather}
The total increment $\Delta n_i$ follows a Skellam distribution. The diffusion approximation for this system reveals the qualitatively different parameter regimes. When we let $K\to\infty$ while fixing the combinations $\gamma\equiv Kg$, $\sigma\equiv Ks$, $\mu\equiv K\mu_0$ and $\phi_i = K f_i$ and setting $t=\tau/K$, the system is described by a Fokker-Planck evolution equation~\cite{Kampen:1992} for the distribution $P(x_s,x_r,t)$ with $x_i\equiv n_i/K$ (see SI text). This scaling exercise is mainly important because it allows to relate systems with small $K$ (100s to 1000s, as necessarily used in numerical analysis) to systems with large $K$ ($\gtrsim10^{8}$, as present in real cancers). It is important to note that in this limit the details of the microscopic model are not important. For example, the effects of selection or carrying capacity could be included in the death rates, without changing the qualitative aspects of the model.

\subsection{Optimal cancer therapy with perfect monitoring}

With the minimal model of drug resistance in cancer introduced above we can start the program of stochastic optimal control to compute  adaptive therapy protocols. The first task is to define the goal of such a program: what is the quantity one aims to maximize? One candidate is the total tumor population size $N$, the long-term reduction of which is the goal of standard therapy~\cite{Gatenby:2009b}. Another very important objective is to maximize the (expected) time until the cancer proceeds to the next, possibly lethal stage. This could mean the emergence of a new cell type with a much higher carrying capacity, e.g. with metastatic potential. We will denote this critical event simply with a `driver' event or `metastasis'. The rate of metastasis emergence is a combination of tumor size and the rate $\nu_0$ (per cell and generation) for the necessary features to appear via mutation.\\
Earlier, the optimal control for the Wright-Fisher evolution example  turned out to be a piecewise constant function of allele frequency. Here, we need to find a control \emph{profile} $\bar{u}(n_s,n_r)$. With perfect information, we would know $n_s$ and $n_r$ at all times and would base the control decision adaptively on these measurements.  As in eq.~\ref{Eq:u1}, the control objective can be expressed as 
\begin{align}
\bar{u}_{0:T}(n_{s0},n_{r0}) &= \argmax_{u_{0:T}}\ \exp\pbr{-\nu_0 \sum_{t=0}^T \pbr{n_{st}+n_{rt}}},
\label{Eq:OptCMetaFree}
\end{align}
where the right hand side is the probability that metastasis has not yet happened by time $T$ generations. Here, the control objective is a non-linear function of the entire trajectory $(n_{s,0:T},n_{r,0:T})$. As such, it is the simplest manifestation of a so-called \emph{risk-sensitive} control problem~\cite{Kappen:2005,Bielecki:1999,Bauerle:2013}. The above formulation assumes a finite (receding) horizon time $T$ and also that control itself is cost-free. In cancer therapy, especially chemotherapy, this is certainly not the case: the side effects of treatment incur a considerable cost in terms of life-quality and medical care. The difficulty, however, lies in quantifying these control costs in a manner that would make them comparable to the potential costs considered here. This important aspect is beyond the scope of this study. 

\begin{figure}
\begin{center}
\includegraphics[width=8cm]{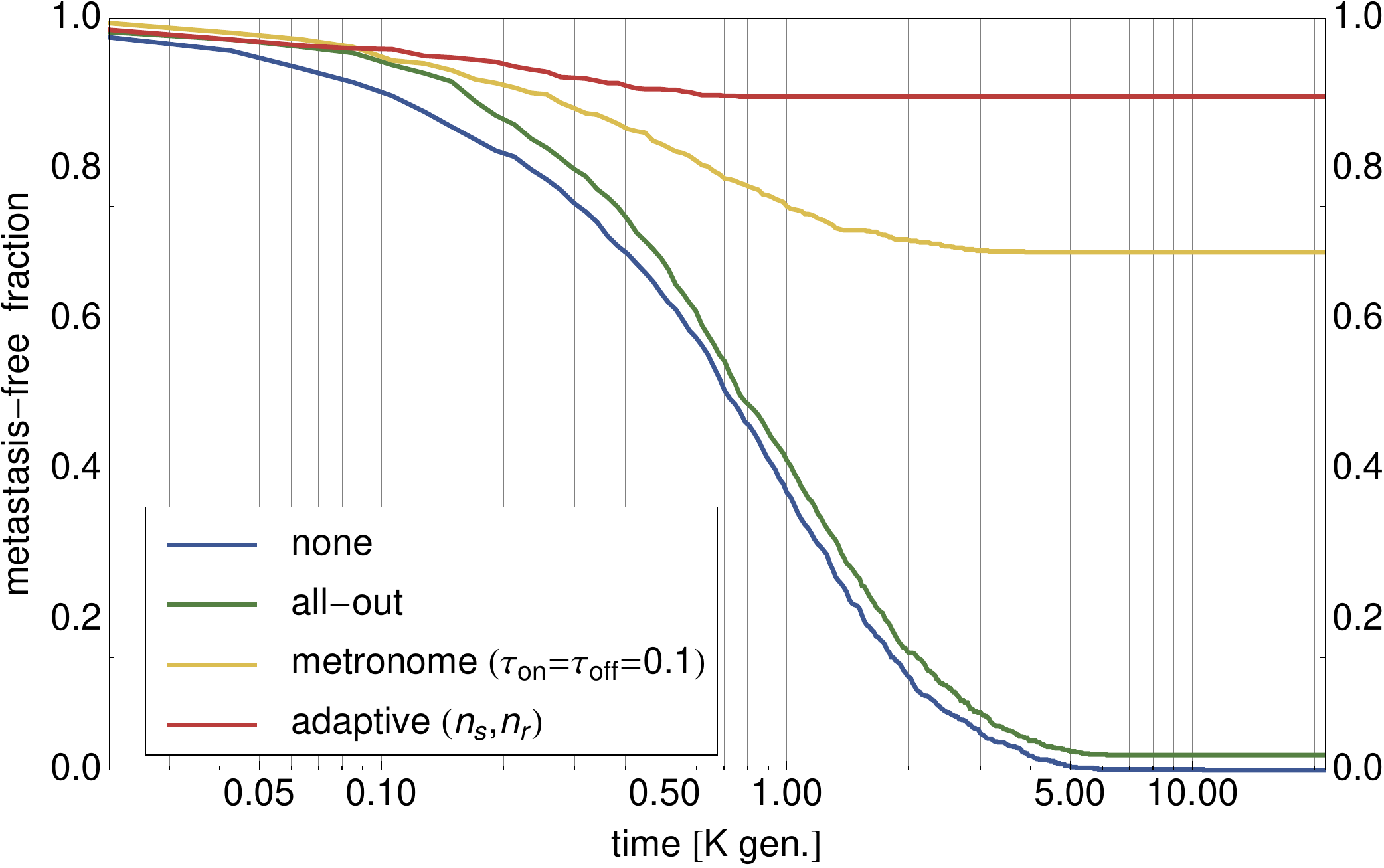}
\caption{Comparison of cancer therapies. 
For the parameter setting of Fig.~2D, different therapies are compared via 1000 forward simulations with $K=10^4$. Shown is the fraction of runs that have not yet developed a metastasis mutation by time $\tau$. All-out maximum dosage therapy ($u\equiv1$) is only slightly better than no therapy ($u\equiv0$) in avoiding metastasis. Much better is metronome therapy with $\tau_{\rm on}=\tau_{\rm off}=0.1\ K\ {\rm gen.}$ with almost 70\% success rate (values numerically optimized). Adaptive therapy removes the tumor in close to 90\% of runs.}
\label{Fig:Therapies}
\end{center}
\end{figure}

The recurrence equation for the cost-to-go $J(n_{s},n_{r},t)$ for the control objective above is given by~\cite{Bauerle:2013}
\begin{align}
J(n_s,n_r,t) &= e^{-\nu_0(n_s+n_r)}  \max_{u\in\cbr{0,1}} \mean{J(t+1;\,u)}
\label{Eq:C2GoCancer}\\
\mean{J(t';u)} &\equiv \sum_{n_s', n_r'} J(n_s',n_r',t')\ W(n_s',n_r' \mid n_s,n_r;\, u)
\label{Eq:MeanC2Go}
\end{align}
with boundary condition $J(T) = 1$.  The (microscopic) transition matrix $W$ is the product of the two Skellam distributions resulting from eq.~\ref{Eq:BD2} (including boundary conditions). With this equation, we  can solve the dynamic programming task numerically for moderate values of $K$. For the numerical analysis, we have to introduce an upper bound  $\tilde{N} \gg K + \sqrt{K}$ for the population size. The resulting control profiles for a number of different parameter regimes is shown in Figure~\ref{Fig:TumorControl}.\\
In the case of $\phi_r=0$, resistant cells are unaffected by the drug. If maintenance of the resistance mechanism is costly ($\sigma>0$), the only way that they can be removed from the population is when selection can act against them. This only happens at $N\sim K$ and with $u=0$ (no drug). If this can take place before the next driver typically appears (if $\sigma = Ks \gg KN \nu_0 \sim K^2\nu_0 \equiv \nu$), then the optimal control protocol is to postpone treatment until the resistant cells are sufficiently cleared from the system (see Fig.~\ref{Fig:TumorControl}A). However, this parameter regime of very high selection against resistance and/or very low rate of driver mutation, and therefore this therapy option, is not  realistic for cancer. For higher values of $\nu / \sigma$, the optimal strategy is to apply the drug earlier (see Fig.~\ref{Fig:TumorControl}B). This procedure can lead to cycles of tumor size reduction followed by regrowth, with the overall effect of extending the time until metastasis.\\
If $\phi_r > \gamma$ (and $\phi_s>\gamma+\sigma$), resistant cells are actually \emph{drug-addicted} and thrive only in its presence. Such a situation would be easy to control with perfect information about $n_s$ and $n_r$. For example, if mutation \emph{between} cell types is rapid ($\mu\gg1$), a majority-rule is optimal ($\bar{u}(n_s>n_r) = 1$ and $\bar{u}(n_r>n_s)=0$ in a fully symmetric setting, see Fig.~\ref{Fig:TumorControl}C). For a lower mutation rate, the optimal profile first tries to amplify one cell type before switching to an environment that is now deadly for most cells present (see Fig.~\ref{Fig:TumorControl}D).\\
The effectiveness of different therapy protocols is compared in Figure~\ref{Fig:Therapies} with 1000 stochastic forward simulations (with $K=10^4$) for the parameter setting of Fig.~\ref{Fig:TumorControl}D. While no therapy ($u=0$) and all-out therapy ($u=1$) both ultimately end with the occurrence of metastasis,  adaptive therapy can bring the tumor size down to zero in the majority of cases. In metronome therapy, the drug is applied (withheld) for fixed time intervals $\tau_{\rm on}$ ($\tau_{\rm off}$). With numerically optimized values of time intervals, metronome therapy is quite competitive.\\
All these  control strategies require perfect information, not only in the sense of the earlier Wright-Fisher example (continuous, synchronous and exact), but also in terms of the inner tumor composition $N=n_s+n_r$, which presupposes that sensitive and resistant cells can be distinguished.

\subsection{Loss of therapy efficacy due to low-resolution monitoring}

\begin{figure*}
\begin{center}
\makebox[\textwidth][c]{
\includegraphics[width=17.6cm]{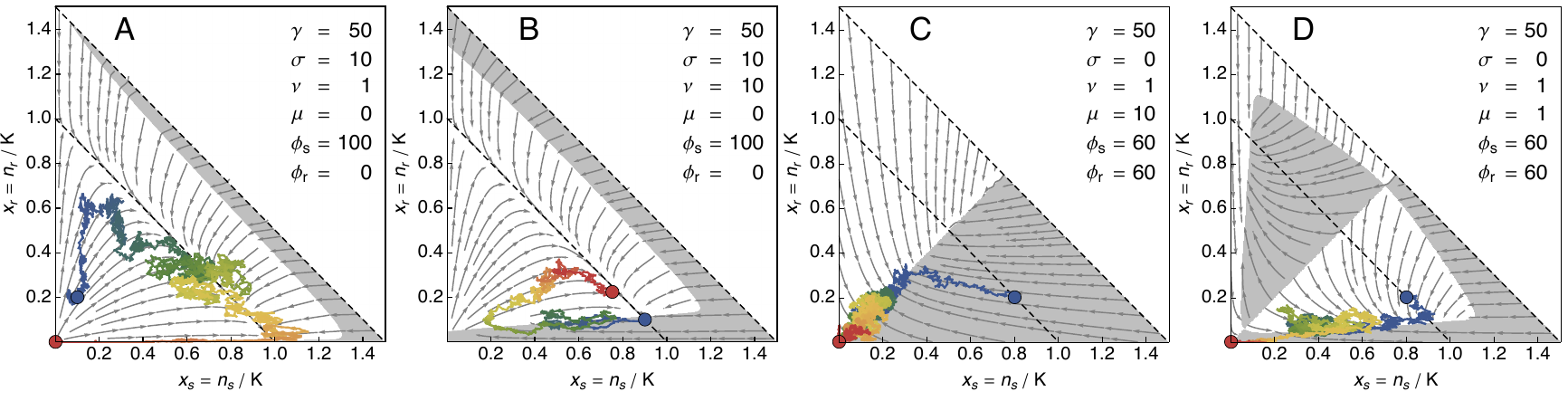}
}
\caption{Control of a tumor cell population. (A-D) The optimal control profile under perfect information about $n_s$ and $n_r$ for different parameters of the cancer model. In the white areas, $\bar{u}=0$ (no drug), whereas in the gray areas, $\bar{u}=1$ (with drug). The arrows indicate the deterministic flow. All profiles were calculated via eq.~\ref{Eq:C2GoCancer} with $T=K/\nu$ gen. and $K=500$ with an absorbing boundary at $N=750$. The sample trajectories were simulated with $K=10^4$ and controlled according to these profiles. The coloring of the trajectories shows the temporal evolution from blue to red. (A) When selection against resistance is stronger than driver emergence, $\sigma \gg \nu$, the optimal protocol is to wait until resistant cells are cleared from the system before the drug is applied. (B) For higher driver emergence rates, the drug is applied earlier, which can lead to cycles. (C) For drug-sensitive ($\phi_s\gg \gamma$) and drug-addicted cells ($\phi_r\gg \gamma$) with high mutation ($\mu\gg1$), the control in the symmetric case ($\phi_s=\phi_r$) is a simple majority rule and very effective. (D) For smaller mutation ($\mu=1$), the optimal strategy first homogenizes the tumor before trying to remove it.}
\label{Fig:TumorControl}
\end{center}
\end{figure*}

There are very few cases where the genetic basis for a drug-resistance mechanism is known and can be specifically monitored~\cite{Thakur:2013,Holohan:2013}. In most cases the regrowth of the tumor under the drug is observed without understanding the exact biological processes responsible for the resistance. Here we aim to find rational control strategies when only the total tumor cell population size can be monitored. The adaptive therapy protocol that was applied by Gatenby et. al in~\cite{Gatenby:2009} (coupling the drug concentration to the tumor size) is one example of such a strategy.\\
Consider the situation where only the total population size $N=n_s+n_r$ can be (perfectly) monitored, while the dynamical laws in eqs.~\ref{Eq:BD1}-\ref{Eq:BD2} and all parameter values are known. Under these circumstances, the perfect-information optimal control profiles from the last section cannot be used directly. 
However, there is still valuable information available. The response $N_{\tau}\to N_{\tau+\Delta\tau}$ of the tumor size to a control choice over a time interval $\Delta\tau$ can give an indication of the inner tumor composition. As we have seen earlier, the length of time interval $\Delta\tau$ should be shorter than all other intrinsic time scales to enable control.
One plausible way to use this information is to continuously update a (posterior) distribution $P(n_s,n_r\mid N_{0:\tau}, u_{0:\tau-\Delta\tau})$ and use it, together with $\bar{u}(n_s,n_r)$,  to determine the next control $u_{\tau}$ as the one that is `correct' in a majority of cases.\\
An entirely different possibility is to first derive an effective propagator $W_{\rm eff}(N_{\tau+\Delta\tau} \mid N_{\tau}, N_{\tau-\Delta\tau}, u_{\tau-\Delta\tau}; u_{\tau})$ and then repeat the cost-to-go calculation of eqs.~\ref{Eq:C2GoCancer}~and~\ref{Eq:MeanC2Go}.
This propagator takes into account not only the current size $N_{\tau}$, but also the last measurement $N_{\tau-\Delta\tau}$ and the last  control decision $u_{\tau-\Delta\tau}$. It follows from the microscopic $W$ used in eq.~\ref{Eq:MeanC2Go} by integrating over the internal degrees of freedom $n_s$ and $n_r$ at the three time points (see SI text). Accordingly, the control profile is now a function of $(N_{\tau-\Delta\tau},u_{\tau-\Delta\tau},N_{\tau})$. For the parameter values leading to the majority-rule in Fig.~\ref{Fig:TumorControl}C, the new control profile is shown in Fig.~\ref{Fig:NuN}. The drug regimen ($u=0$ or $1$) is maintained as long as the tumor size decreases sufficiently. At the first sign of possible reversal, the regimen is switched.

\section{Discussion}
We used stochastic control theory to quantify optimal control strategies for models of evolving populations. We further demonstrated how control can be maintained with finite resources, when the monitoring necessary for adaptive control is imperfect. These strategies all depend on our ability to \emph{anticipate} evolution, i.e. on a knowledge of the relevant equations of motion and their parameter values. For cancer, such detailed knowledge of evolutionary dynamics is certainly not yet available.  Sequencing technologies are facing up to the challenge of tumor control with finite information, already accelerating progress in monitoring of serial biopsies of tumors, circulating tumor cells or cell-free tumor DNA in the bloodstream~\cite{Schuh:2012,Murtaza:2013}. Once such time-resolved data become prevalent, we can start to learn and improve dynamical tumor models and compute their optimal control strategies. For instance, genetic heterogeneity within the tumor is now becoming quantifiable from sequencing data via computational inference~\cite{Oesper:2013,Roth:2014,Fischer:2014}.  Heterogeneity and subclonal dynamics have been found to have an impact on treatment strategy selection~\cite{Beckman:2012p5337}. Furthermore, all other available sources of clinical data, such as medical imaging, can provide additional high-resolution information and should be integrated into a truly personalized and data-driven tumor-control effort (see e.g.~\cite{Haeno:2012p5217} for imaging data based computational modelling of pancreatic cancer growth dynamics to guide treatment choice and ~\cite{Yuan:2012} for integrative analysis of imaging and genetic data). \\
Beyond cancer, the need to control evolving populations is a key global health challenge as resistant strains of bacteria, viruses and parasites are spreading~\cite{zurWiesch:2011p4914,Goldberg:2012p4885,Greene:2013p5390}. Similarly, pest resistance is also posing a danger to food supplies and needs to be contained. Any long term success in controlling evolution depends, at the very least, on mastering the following components. Firstly, on a quantitative understanding of the underlying evolutionary dynamics. Progress in understanding is best demonstrated by predicting evolution; this has so far proven difficult, even in the short term. Nevertheless, new population genetic approaches applied to data are promising -- see influenza strain prediction in Ref.~\cite{uksza:2014p5387}.
Secondly, the success of control will depend on the availability of a sufficient arsenal of non-cross resistant therapeutic agents. These therapeutics should be combined with the ability to decide an appropriate drug regimen given the genetic and phenotypic structure of the population. Large-scale drug vs. cell line screens are systematically pushing this component forward (see e.g.~\cite{Garnett:2012p4901}). And finally, on the ability to monitor the evolution of target population and act rationally based on this information; the topic of this paper.

\flushleft {\bf Acknowledgments}\\

We would like to acknowledge the Wellcome Trust for support under grant references 098051 and 097678. AF is in part supported by the German Research Foundation (DFG) under grant number FI 1882/1-1. We would like to thank C. Illingworth for discussions and J. Berg, C. Callan, C. Greenman and P. Van Loo for comments on an earlier version of the manuscript.
 
\bibliographystyle{pnas-bolker}
\bibliography{Monitoring}

\renewcommand\thefigure{S\arabic{figure}}    
{\flushleft\LARGE
\textbf{Supporting Text:}\\[2ex]
\Large
\textbf{The value of monitoring to control evolving populations}
}

{\flushleft
Andrej Fischer$^{1}$,
Ignacio V\'{a}zquez-Garc\'{i}a$^{1,2}$ and
Ville Mustonen$^1$}

{\flushleft
${}^1$Wellcome Trust Sanger Institute, Hinxton, Cambridge CB10 1SA, UK\\[1ex]
${}^2$Department of Applied Mathematics and Theoretical Physics, University of Cambridge, Cambridge CB3 0WA, UK
}

\normalsize

\section{Bang-bang control of Wright-Fisher evolution}
Consider the Wright-Fisher model of evolution of a bi-allelic population of finite size $N\gg 1$ with (symmetric) mutation rate $\mu=N\mu_0\geq 0$ and (intrinsic) selection coefficient $\sigma=N s>0$. In the diffusion approximation~\cite{Kampen:1992,Gardiner:2009}, the probability distribution $P(x,t)$ for the allele frequency $x\equiv n_A/N$ of the $A$-allele changes in time according to the following Fokker-Planck equation~\cite{Ewens:2004},
\begin{gather}
\partial_t P(x,t)= \bbr{-\partial_x \pbr{\sigma\,x (1-x)+\mu(1-2x)} + \tfrac{1}{2}\partial_x^2 x(1-x)}\ P(x,t)
\label{Eq:WFFP}
\end{gather}
with boundary condition $P(x,0)=\delta(x-x_0)$. In the limit $\mu\to0$, consider now the control task of maintaining an initial polymorphism $0<x_0<1$ for as long as possible by linearly changing the selection coefficient instantaneously in response to and as a function of $x_t$:
\begin{gather}
\sigma \to \sigma + u(x_t),\quad u \in \bbr{-u_c,0},\ u_c > \sigma.
\end{gather}
The optimal control strategy $\bar{u}(x)$ will maximize the average survival time of the polymorphism,
\begin{gather}
\bar{u} = \argmax_{u} \mean{T}_{x_0}
\end{gather}
where $\mean{\ldots}_{x_0}$ is the average over trajectories starting at $x_0$ and the maximization is over all functions $u:\bbr{0,1}\to\bbr{-u_c,0},\ x\mapsto u(x)$. In this setting, using control itself does not incur a cost and does not enter the maximization objective. It can be shown~\cite{Kappen:2005}, that maximization is then to be carried out  in the smaller function space $u:\bbr{0,1}\to\cbr{-u_c,0}$, where only the extremal  control values are used. This type of control is called \emph{bang-bang}. It is also evident that the optimal control strategy, maximizing the mean survival time, will be a piecewise constant function with a single step at a threshold $x_c\in\pbr{0,1}$.
\begin{gather}
\bar{u}(x) \equiv \begin{cases}
0, &x<x_c\\
-u_c,  &x\ge x_c 
\end{cases}
\end{gather}
If both the intrinsic selection coefficient $\sigma$, and the control strength $u_c$ are given, then we need to optimize only a single parameter, the threshold $x_c$. 

\subsection{Analytical evaluation of the mean first passage time}
\begin{figure}[ht]
\begin{center}
\includegraphics[width=0.7\textwidth]{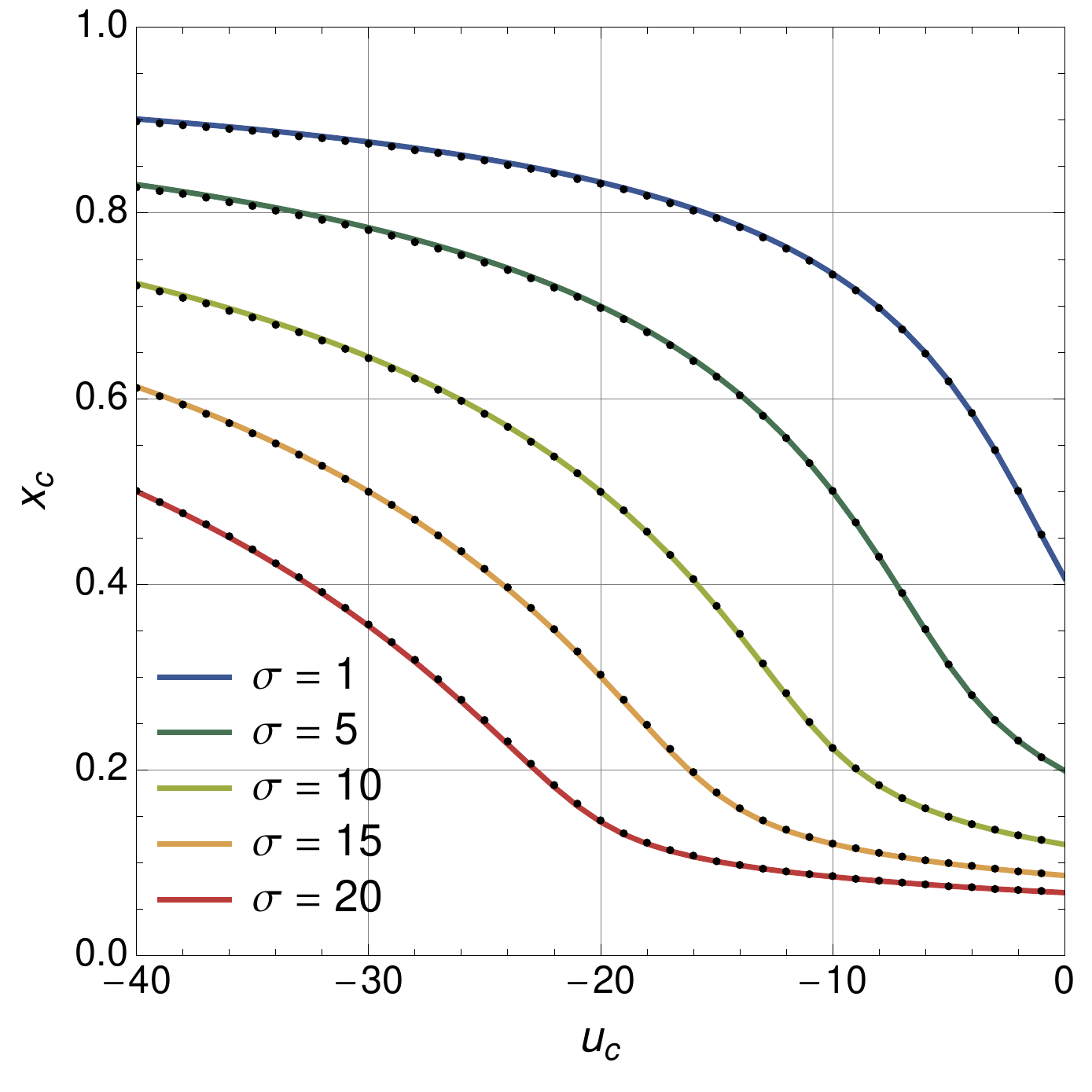}
\caption{\small The control-switch frequency $x_c(u_c,\sigma)$ found by maximizing the analytical expression for the mean survival time (solid lines) and by backwards iteration of eq.~\ref{Eq:C2G1} in the main text. Note that $x_c(-2\sigma,\sigma)=0.5$.}
\label{Fig:xc}
\end{center}
\end{figure}

Under bang-bang control, the effective selection coefficient $\sigma + \bar{u}$ is frequency dependent but still piecewise constant. For $\sigma > 0$ and $\sigma +  u_c < 0$, the population experiences an upward drift for $x<x_c$ and a downward drift for $x>x_c$. If the drift forces in both domains are strong ($\sigma\gg1$ and $\sigma + u_c \ll -1$), then a typical population \emph{that is still polymorphic} will most likely be in the vicinity of $x_c$ at any one point in time (see also Figure~\ref{Fig:TumorControl}A in the main text). One can then try to calculate the mean first passage time for trajectories starting at $x_c$. The formula can be found using standard theory of stochastic processes~\cite{Gardiner:2009}. Let us momentarily re-insert an arbitrary initial frequency $x_0$,
\begin{gather}
\mean{T}_{x_0} = 
\frac{
\pbr{\int_0^{x_0} \frac{\d y}{\psi(y)}} 
\int_{x_0}^1 \frac{\d y'}{\psi(y')} \int_0^{y'} \frac{\d z\ \psi(z)}{z(1-z)}
- 
\pbr{\int_{x_0}^1 \frac{\d y}{\psi(y)}} 
\int_0^{x_0} \frac{\d y'}{\psi(y')} \int_0^{y'} \frac{\d z\ \psi(z)}{z(1-z)}
}{
\frac{1}{2}\ \int_0^1 \frac{\d y}{\psi(y)}
}\\[1ex]
{\rm with}\quad \psi(z) \equiv \exp\bbr{ 2\sigma z + 2 u_c\,(z-x_c)\, \Theta(z-x_c)}.
\end{gather}
The mean first passage time depends on the initial point $x_0$, which can be either below or above $x_c$. This will affect all the integrals, so let us write
\begin{gather}
\mean{T}_{x_0} = \mean{T}_{x_0}^+\ \Theta(x_0-x_c) + \mean{T}_{x_0}^-\ \Theta(x_c-x_0) \\[1ex]
{\rm with}\quad \mean{T}_{x_0}^{\pm} = \frac{ I^{\pm}_{0,x_0}\ J^{\pm}_{x_0,1} - I^{\pm}_{x_0,1}\ J^{\pm}_{0,x_0}}{\frac{1}{2}\ I_{0,1}}.
\end{gather}
The integrals in this ratio can now be computed one by one. They all have analytical solutions. 
\begin{align}
I_{0,x_0}&\equiv \int_0^{x_0} \frac{\d y}{\psi(y)} = I_{0,x_0}^+\Theta(x_0-x_c) + I_{0,x_0}^-\ \Theta(x_c-x_0) \\
I_{0,x_0}^- &= \frac{1}{2\sigma}\pbr{1 - e^{-2\sigma x_0}}\\
I_{0,x_0}^+ &=  \frac{1}{2\sigma}\pbr{1 - e^{-2\sigma x_c}} + \frac{e^{2u_c x_c}}{2 (\sigma + u_c)}\, \pbr{e^{-2(\sigma+u_c) x_c} - e^{-2(\sigma+u_c) x_0} }
\end{align}
\begin{align}
I_{x_0,1}&\equiv \int_{x_0}^1 \frac{\d y}{\psi(y)} = I_{x_0,1}^+\Theta(x_0-x_c) + I_{x_0,1}^-\ \Theta(x_c-x_0) \\
I_{x_0,1}^- &=  \frac{1}{2\sigma}\pbr{e^{-2\sigma x_0} - e^{-2\sigma x_c}} + \frac{e^{2u_c x_c}}{2 (\sigma + u_c)}\, \pbr{e^{-2(\sigma+u_c) x_c} - e^{-2(\sigma+u_c)} }\\
I_{x_0,1}^+ &= \frac{e^{2u_c x_c}}{2 (\sigma + u_c)}\, \pbr{e^{-2(\sigma+u_c) x_0} - e^{-2(\sigma+u_c)} }
\end{align}
%
\begin{align}
J_{x_0,1}&\equiv \int_{x_0}^1 \frac{\d y'}{\psi(y')} \int_0^{y'} \frac{\d z\ \psi(z)}{z(1-z)}
 = J_{x_0,1}^+\Theta(x_0-x_c) + J_{x_0,1}^-\ \Theta(x_c-x_0) \\
J_{x_0,1}^- &=  \pbr{G(x_0,\sigma) - G(x_c,\sigma)} - \frac{F(0,\sigma)}{2\sigma}\pbr{e^{-2\sigma x_0} - e^{-2\sigma x_c}} \\
&\phantom{=}  + e^{2u_c x_c}\,\pbr{G(x_c,\sigma+u_c) - G(1,\sigma+u_c)} \non \\
&\phantom{=} + \frac{e^{2u_c x_c}}{2(\sigma + u_c)}\,\pbr{e^{-2(\sigma + u_c) x_c} - e^{-2(\sigma + u_c)}}\, \pbr{F(x_c,\sigma) - F(0,\sigma) - F(x_c,\sigma + u_c)}\non \\
J_{x_0,1}^+ &= \frac{e^{2u_c x_c}}{2(\sigma +u_c)}\pbr{e^{-2(\sigma+u_c)x_0}  - e^{-2(\sigma+u_c)}} \, \pbr{F(x_c,\sigma) - F(0,\sigma) - F(x_c,\sigma + u_c)}\non\\
&\phantom{=} + e^{2u_c x_c}\ \pbr{G(x_0,\sigma+u_c) - G(1,\sigma+u_c)} 
\end{align}
%
\begin{align}
J_{0,x_0}&\equiv \int_{0}^{x_0} \frac{\d y'}{\psi(y')} \int_0^{y'} \frac{\d z\ \psi(z)}{z(1-z)}
 = J_{0,x_0}^+\Theta(x_0-x_c) + J_{0,x_0}^-\ \Theta(x_c-x_0) \\
J_{0,x_0}^- &=  \pbr{G(0,\sigma) - G(x_0,\sigma)} - \frac{F(0,\sigma)}{2\sigma}\pbr{1-e^{-2\sigma x_0} } \\
J_{0,x_0}^+ &= \pbr{G(0,\sigma) - G(x_c,\sigma)} - \frac{F(0,\sigma)}{2\sigma}\pbr{1-e^{-2\sigma x_c} } \\
&\phantom{=}  + e^{2u_c x_c} \pbr{G(x_c,\sigma+u_c) - G(x_0,\sigma+u_c)} \non\\
&\phantom{=} + \frac{e^{2u_c x_c}}{2(\sigma +u_c)}\pbr{e^{-2(\sigma+u_c)x_c}  - e^{-2(\sigma+u_c)x_0}} \, \pbr{F(x_c,\sigma) - F(0,\sigma) - F(x_c,\sigma + u_c)}\non
\end{align}
The solutions include the following functions
\begin{align}
F(x,\sigma)\ &\equiv\ \ExpEi(2\sigma x) - e^{2\sigma} \ExpEi(-2\sigma (1-x)),\\
G(x,\sigma)\ &\equiv\ \frac{e^{-2\sigma x}}{2\sigma} F(x,\sigma) + \frac{1}{2\sigma} \log\pbr{\frac{1-x}{x}}\\
 \ExpEi(z)\ &\equiv\ -\int_{-z}^{\infty}\d t\ \frac{e^{-t}}{t}
\end{align}
where we also used the following identities,
\begin{gather}
F(x,\sigma) = - e^{-2\sigma} F(1-x,-\sigma)\quad {\rm and}\quad G(x,\sigma) = G(1-x,\sigma).
\end{gather}
Finally, $\mean{T}_{x_0=x_c}$ can be evaluated numerically and maximized with respect to $x_c$ to find this critical control threshold. The result is shown in Figure~\ref{Fig:xc} and is compared to the corresponding result of the cost-to-go backwards iteration for the discrete system.

\begin{figure}[ht]
\begin{center}
\includegraphics[width=0.8\textwidth]{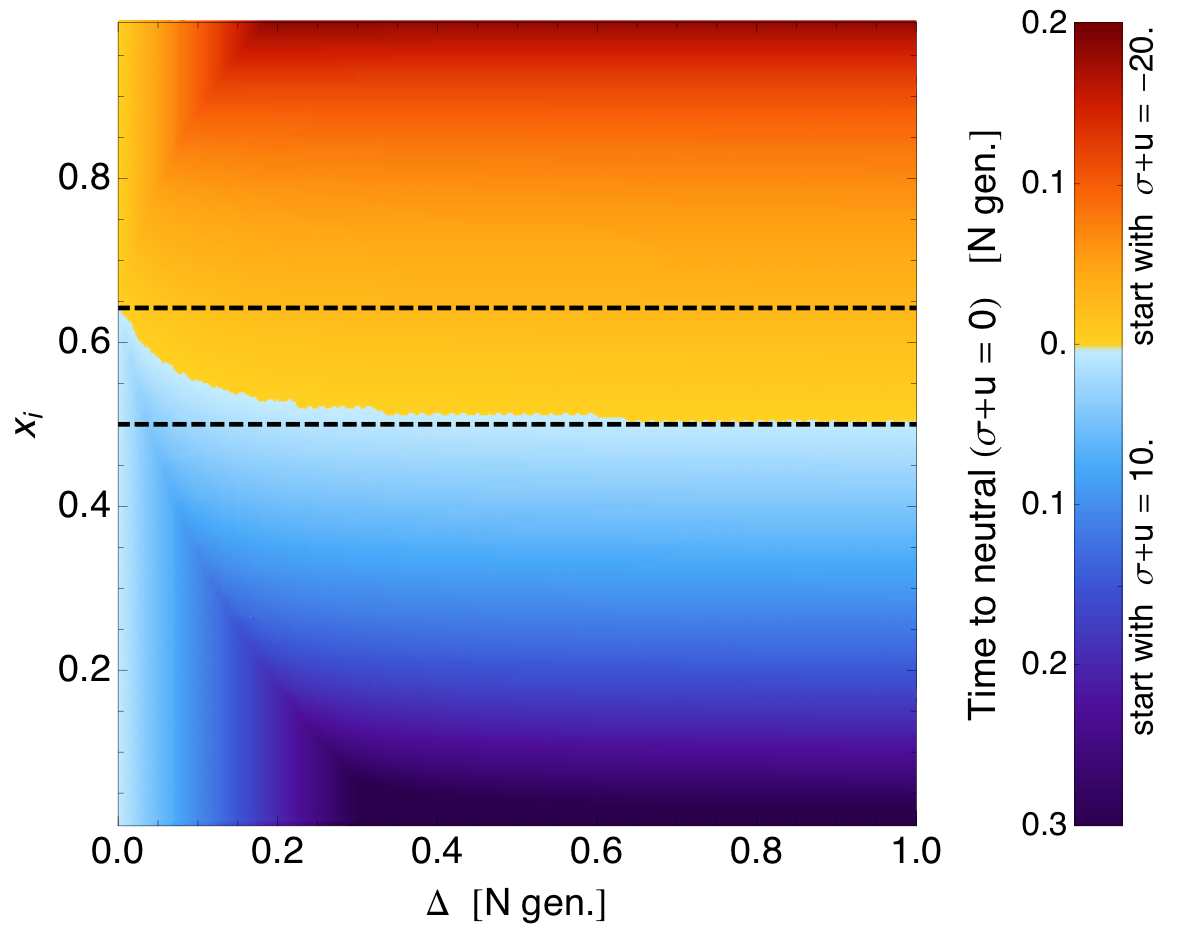}
\end{center}
\caption{\small For finite time $\Delta$ between consecutive measurements, the pre-emptive control aims for a safe position $x_{\rm safe}$ away from the boundaries (boundary between blue and orange) by switching to a neutral regime ($u=-\sigma$) after a certain waiting time (coloring, see legend). At $x_{\rm safe}$, the waiting time to neutral is zero, i.e. the system is \emph{immediately} set to neutral. As $\Delta$ becomes bigger, $x_{\rm safe}$ moves from $x_c\approx 0.644$ to $0.5$ and the control strategy shifts from \emph{playing-to-win} to \emph{playing-not-to-lose}.}
\label{Fig:win-not-lose}
\end{figure}

\section{Minimal model of drug resistance in cancer}
The qualitative aspects of the minimal cancer model introduced in the main text can be analyzed using a system size expansion~\cite{Kampen:1992}, with the carrying capacity $K$ as a large parameter. The expansion entails the parameter scaling
\begin{gather}
K\to\infty\quad{\rm with}\quad \gamma \equiv K g,\ \sigma \equiv Ks,\ \mu \equiv K\mu_0,\ \phi_{s,r}\equiv K f_{s,r}\quad  {\rm const.}
\end{gather}
together with a scaling of time via $\tau = t/N$ (with $t$ measured in generations, i.e. Poisson population updates). The typical relative scale of the model parameters is
\begin{gather}
K \gg \phi > \gamma\ \gg\ \sigma,\  \mu \geq 0.
\end{gather}
The expansion of birth and death rates in $K$ is as follows:
\begin{align}
B_s(x_s,x_r) &= \frac{(K+\gamma+\sigma)\, K\,x_s}{K + \gamma\, (x_s+x_r) + \sigma\, x_s} + \mu (x_r-x_s) \non \\[1ex]
 &= K\, x_s + \gamma\,x_s\,(1-x_s-x_r)+\sigma\, x_s\,(1-x_s) + \mu(x_r-x_s)+\order{K^{-1}}\non\\[1ex]
 &\equiv K\,x_s + b_s(x_s,x_r) + \order{K^{-1}}\\[1ex]
 B_r(x_r,x_s) &= K\,x_r + \gamma\, x_r\,(1-x_s-x_r) -\sigma\, x_s\, x_r + \mu(x_s-x_r)+ \order{K^{-1}}\non\\
  &\equiv K\,x_r + b_r(x_s,x_r) + \order{K^{-1}}\\[1ex]
  D_s(x_r,x_s) &= K\,x_s + u\,\phi_s\,x_s,\quad
  D_r(x_r,x_s) = K\,x_r + (1-u)\,\phi_r\,x_r
\end{align}
The differential growth rate $\sigma$ and the drug-related death rates $\phi_{s,r}$ break the symmetry of the model, such that there is no closed growth law for the total population size $N = n_s+n_r$ alone: even ignoring boundary terms (at $n_s=0$ and $n_r=0$) the tumor size would evolve according to
\begin{align}
\Delta N &\sim {\rm Skellam}(B_s+B_r, D_s + D_r)\\[1ex]
\mean{\Delta N} &= B_s + B_r - D_s - D_r = b_s+b_r-d_s-d_r\non\\
&= \gamma\, x(1-x) + \sigma\, x_s\, (1-x) - u\,\phi_s\,x_s - (1-u)\phi_r\,x_r + \order{K^{-1}}
\end{align}
with $x \equiv x_s+x_r = N/K$. The role of $K$ as carrying capacity (for $u=0,\ \phi_r=0$) is now apparent via $\mean{\Delta{N}}(x=1)=0$. The Fokker-Planck equation for this model in the variables $(x_s,x_r)$ is given by
\begin{gather}
\partial_{\tau}P(x_s,x_r,\tau) = \bbr{-\partial_{x_s}(b_s-d_s) - \partial_{x_r}(b_r-d_r) + \pbr{\partial_{x_s}^2 + \partial_{x_r}^2}(x_s+x_r)} P(x_s,x_r,\tau)
\end{gather}
The form of the birth and death rates above suggests a transformation of variables.
\begin{gather}
(x_s,x_r) \to \pbr{x\equiv x_s+x_r,\ y\equiv \frac{x_s}{x_s+x_r}}\ \Rightarrow \ \pbr{x_s = x\,y,\ x_r = x(1-y)}
\end{gather}
The time  evolution of the mean values of these new variables is now given by~\cite{Kampen:1992}
\begin{align}
\partial_{\tau}\mean{x} &= \mean{b_s+b_r-d_s-d_r}\non \\[1ex]
&= \mean{(\gamma+\sigma y)x(1-x) - x(u\phi_s y + (1-u)\phi_r(1-y))}\\[1ex]
\partial_{\tau}\mean{y} &= \mean{\frac{1-y}{x}(b_s-d_s)-\frac{y}{x}(b_r-d_r)} \non\\
&= \mean{\pbr{\sigma-u\phi_s+(1-u)\phi_r}\, y(1-y) + \mu(1-2y)}
\end{align}
The evolution of the mean relative fraction $\mean{y}$ of sensitive cells is equivalent to the evolution of the mean value of the polymorphism frequency within the controlled one-locus two-alleles Wright-Fisher model discussed earlier (see eq.~\ref{Eq:WFFP}).

\subsection{Numerical test of the cost-to-go calculation}
\begin{figure}[ht]
\begin{center}
\includegraphics[width=0.7\textwidth]{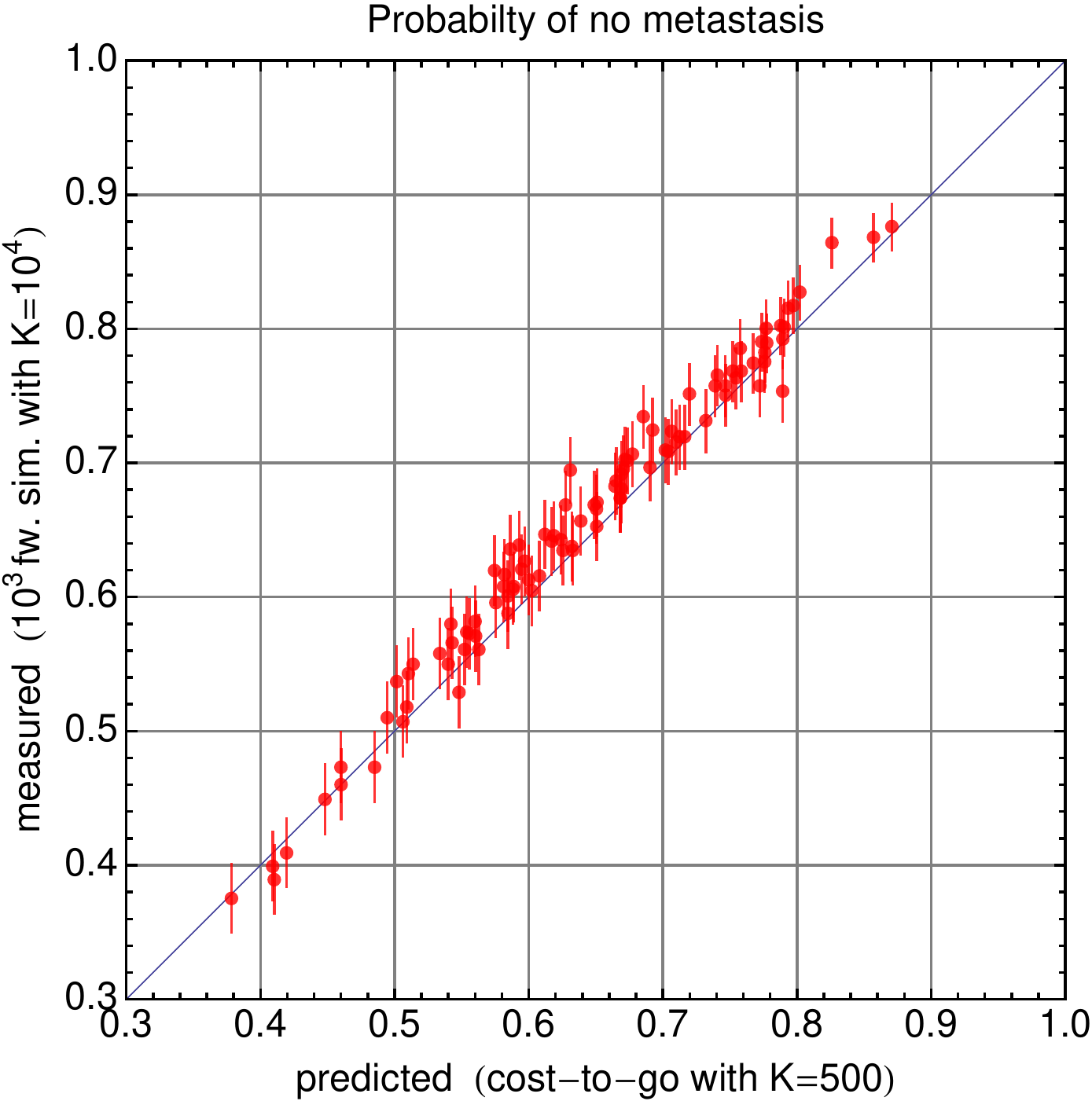}
\caption{\small Comparison of the predicted probability that metastasis has not yet occurred by time $T=1/\nu$ in a cancer cell population optimally controlled according to the profile (and parameters) shown in main text Figure~\ref{Fig:TumorControl}A to the measured fraction of $10^4$ forward simulations with that property. The prediction follows from the cost-to-go dynamic programming calculation (see eq.\ref{Eq:C2GoCancer} in the main text) performed numerically with $K=500$ and $N\leq 750$. The forward simulations were carried out with $K=10^4$, $N\leq 1.5 K$.}
\label{Fig:FWtest}
\end{center}
\end{figure}

An optimal control strategy fulfilling eq.~\ref{Eq:OptCMetaFree} in the main text can be found numerically by backwards iteration of the cost-to-go recurrence equation \ref{Eq:C2GoCancer} using the exact discrete propagator $W(n_s',n_r'\mid n_s,n_r)$ defined by the Skellam distribution implicit in eq.~\ref{Eq:BD2}. To test the sanity of the resulting profile $\bar{u}(x_s,x_r)$, we can evaluate the associated cost function directly using a large ensemble of forward simulations. It should be noted that due to memory and time limitations, the backwards iteration can only be performed with a rather small system size of the order $\max(n_s,n_r) \leq 10^3$. For the forward simulations, only the milder time restriction holds, such that $K\sim\order{10^4}$ is possible. To make the two results comparable, it is necessary to use the same \emph{scaled} parameters $\sigma=K s$ etc. In Figure~\ref{Fig:FWtest}, we compare the probability that metastasis has not yet occurred by time $T=1/\nu$ (the control objective to be maximized) as predicted by the cost-to-go calculation with the direct observation of this event in $10^3$ forward simulations with $K=10^4$ in the parameter setting of Figure~\ref{Fig:TumorControl}A in the main text.

\subsection{Uniform expansion of the Skellam distribution}
The probability mass function of the Skellam distribution with parameters $(\mu_1,\mu_2)$ is given by
\begin{gather}
n_1 \sim {\rm Pois}(\mu_1),\ n_2 \sim {\rm Pois}(\mu_2)\
\Rightarrow\ n \equiv n_1-n_2\ \sim\ {\rm Skellam}(\mu_1,\mu_2),\ n\in \mathbb{Z}\non\\[1ex]
{\rm with}\quad {\rm Skellam}(n\mid \mu_1,\mu_2) = e^{-\mu_1-\mu_2}\pbr{\tfrac{\mu_1}{\mu_2}}^{n/2}\,I_{\abs{n}}\pbr{2\sqrt{\mu_1\mu_2}}.
\end{gather}
The modified Bessel function $I_n(z)$ could, in principle, be evaluated for fixed $z$ via the following recurrence relation,
\begin{gather}
I_{n-1}(z) - I_{n+1}(z) = \frac{2n}{z}\ I_{n}(z).
\end{gather}
However, due to a lack of numerical stability of this recurrence, we have here used the uniform expansion of the Bessel function instead~\cite{Olver:1974},
\begin{gather}
I_{\nu}(\nu z)\ \xrightarrow{\nu\to\infty}\ \frac{e^{\nu\,\eta}}{\sqrt{2\pi\nu} \pbr{1+z^2}^{1/4}}\ \pbr{1+\order{\frac{1}{\nu}}}\\[1ex]
{\rm with}\quad \eta \equiv \sqrt{1+z^2} + \ln(z) - \ln\pbr{1+\sqrt{1+z^2}}.
\end{gather}
This expansion has the additional benefit that we can simply use it for the logarithm of the Skellam distribution,
\begin{align}
\log\ {\rm Skellam}\pbr{n} &\approx a+b\,n + \norm{(n,z)} - \tfrac{1}{2}\log\norm{(n,z)} + \abs{n}\log \frac{z}{\abs{n}+\norm{(n,z)}},\\
{\rm with}\quad a &\equiv -(\mu_1+\mu_2) - \tfrac{1}{2}\log(2\pi),\non\\
b &\equiv \tfrac{1}{2}\log\pbr{\tfrac{\mu_1}{\mu_2}},\non\\
 z &\equiv 2\sqrt{\mu_1\mu_2},\non\\
 \norm{(n,z)} &\equiv  \sqrt{n^2+z^2}\non.
\end{align}
The quality of this approximation is also implicit in the simulation test results shown in Figure~\ref{Fig:FWtest}.

\subsection{Control with limited information ($N$ only)}
\begin{figure}[ht]
\begin{center}
\includegraphics[width=\textwidth]{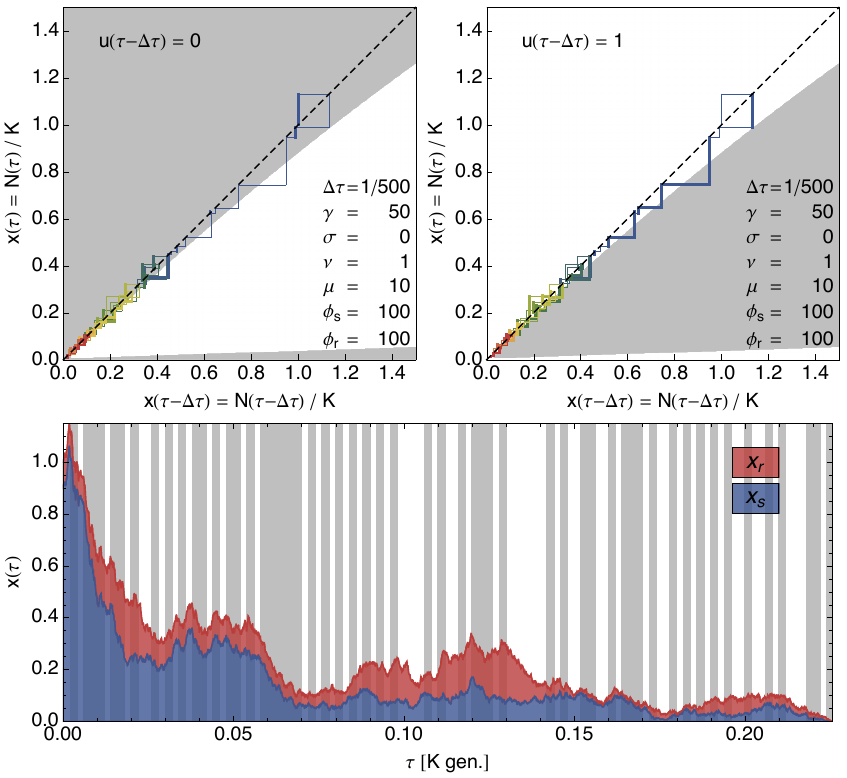}
\end{center}
\caption{Control of a tumor via its total size. 
In the parameter setting of Figure~\ref{Fig:TumorControl}C, a majority rule would be optimal with perfect information. But when only the total population size $N=n_s+n_r$ can be measured, the needed information is not directly available. The control profile above tries to estimate the inner composition of the tumor indirectly from the immediate response $N(\tau-\Delta\tau)\to N(\tau)$ to the presence  ($u(\tau - \Delta \tau)=1$, gray areas) or absence of the drug ($u(\tau - \Delta \tau)=0$, white areas). The trajectory shown in the lower panel is also shown in the control profile above, where thick lines indicate a response big enough to continue the current drug regimen.}
\label{Fig:NuN}
\end{figure}

The optimal control profiles shown in Figure~\ref{Fig:TumorControl} in the main text are only applicable with perfect information of the tumor composition $(n_s,n_r)$. If only the total population size $N$ can be measured, then there is a different strategy to make a control decision: compute the reduced propagator $W(N_{\tau+\Delta\tau}\mid N_{\tau-\Delta\tau},u_{\tau-\Delta\tau},N_{\tau};\ u_{\tau})$ and derive a new control profile that depends on the last two measurements and the last applied control. This is clearly an approximation, such that the resulting control protocol can not be considered optimal in the mathematical sense. In deriving the reduced propagator, we use the shorthand notation $N'=N_{\tau+\Delta\tau}$, $N= N_{\tau}$, $M=N_{\tau-\Delta\tau}$, $u=u_{\tau}$, $v=u_{\tau-\Delta\tau}$ and $n=n_s$.
\begin{align}
W\pbr{ N'\mid N,\ M,\ v;\ u} &=  \sum_{n'=0}^{N'}\ \sum_{n=0}^{N}\ \sum_{m=0}^{M}\ 
 W\pbr{n',\,N'-n' \mid n,\, N - n;\, u}\times\dots
 \label{Eq:EffW}\\
  &\dots\times\ P\pbr{ n,\, N-n\mid m,\, M-m,\, N,\, v}\ P\pbr{m,\, M-m\mid M,\, N,\, v}\non
\end{align}
The first term on the right hand side is the microscopic propagator, expressed as the product of the two Skellam distributions for $n_s$ and $n_r$. The second term is the probability to go from $(m,M-m)$ to $(n,N-n)$ under control $v$, \emph{given} that the final population size is $N$,
\begin{gather}
P\pbr{ n,\, N-n\mid m,\, M-m,\, N,\, v} = \frac{ W\pbr{n,\, N-n\mid m,\, M-m;\ v} }{ \sum_{k=0}^N W\pbr{k,\, N-k\mid m,\, M-m;\ v}}.
\end{gather}
The third and last term is the probability that the system was at $(m,M-m)$, \emph{given} that a transition took place from $M$ to $N$ under control $v$,
\begin{gather}
P\pbr{m,\, M-m\mid M,\, N,\, v} = \frac{ \sum_{n=0}^N W\pbr{n,\, N-n\mid m,\, M-m;\ v} }{\sum_{k=0}^M\, \sum_{n=0}^N W\pbr{n,\, N-n\mid k,\, M-k;\ v}}.
\end{gather}
All these conditional probabilities can be approximated using the logarithmic expansion of the Skellam distribution above. For the parameter setting of Figure~\ref{Fig:TumorControl}C in the main text, the resulting control profile is shown in Figure~\ref{Fig:NuN}.


\end{document}